\documentclass[lettersize,journal]{IEEEtran}
\usepackage{amsmath,amsfonts}
\usepackage{algorithmic}
\usepackage{array}
\usepackage[caption=false,font=normalsize,labelfont=sf,textfont=sf]{subfig}
\usepackage{textcomp}
\usepackage{stfloats}
\usepackage{url}
\usepackage{verbatim}
\usepackage{graphicx}
\def\BibTeX{{\rm B\kern-.05em{\sc i\kern-.025em b}\kern-.08em
    T\kern-.1667em\lower.7ex\hbox{E}\kern-.125emX}}
\usepackage{balance}

\usepackage{algorithmic}
\usepackage{algorithm}
\usepackage[T1]{fontenc}

\begin{document}
\title{Significant improvement of lossy compression rate and speed of HPC data using perceptron parallelized compression}
\author{
    \IEEEauthorblockN{
		Xinzhe Chen\IEEEauthorrefmark{1}, 
		Jianjiang Li\IEEEauthorrefmark{2}
    }

	\IEEEauthorblockA{\IEEEauthorrefmark{1}University of Science and Technology Beijing,Beijing 100083,China}
	
    \IEEEauthorblockA{\IEEEauthorrefmark{2}Department of Computer Science and Technology,University of Science and Technology Beijing,Beijing 100083,China \\ Email: lijianjiang@ustb.edu.cn}
\thanks{\IEEEauthorrefmark{2} Corresponding author}}

%\markboth{Journal of \LaTeX\ Class Files,~Vol.~18, No.~9, September~2020}%
%{Significant improvement of lossy compression rate and compression speed of HPC data using perceptron parallelized compression}

\maketitle

\begin{abstract}
The escalating surge in data generation presents formidable challenges to information technology, necessitating advancements in storage, retrieval, and utilization. With the proliferation of artificial intelligence and big data, the ``Data Age 2025''\cite{ref13} report forecasts an exponential increase in global data production. The escalating data volumes raise concerns about efficient data processing. The paper addresses the predicament of achieving a lower compression ratio while maintaining or surpassing the compression performance of state-of-the-art techniques.

This paper introduces a lossy compression framework grounded in the perceptron model for data prediction, striving for high compression quality. The contributions of this study encompass the introduction of positive and negative factors within the relative-to-absolute domain transformation algorithm, the utilization of a three-layer perceptron for improved predictive accuracy, and data selection rule modifications for parallelized compression within compression blocks. Comparative experiments with SZ2.1's PW\_REL mode demonstrate a maximum compression ratio reduction of $17.78\%$.
    
The article is structured as follows: the introduction highlights the data explosion challenge; related work delves into existing solutions; optimization of mapping algorithms in the relative and absolute domains is expounded in Section 3,the design of the new compression framework is detailed in Section 4,In Section 5 we describe the whole process and give pseudo-code, and in Section 6, our solution is evaluated. Finally, in Section 7, we provide an outlook for future work.
\end{abstract}


\section{Introduction}
\IEEEPARstart{T}{he} exponential growth of information generated by humans poses significant challenges to information technology in terms of storage, retrieval, and utilization. Early computers were primarily employed for scientific computations. However, as computers have evolved and their applications expanded, the volume of data they need to process has increased substantially. This data encompasses diverse types and intricate structures, ranging from simple numeric data to non-numeric and structured data. With the progression of artificial intelligence and big data, the exponential surge in data has become evident. The ``Data Age 2025''\cite{ref13} report by IDC indicates that global data production will rise from 33 ZB in 2018 to 175 ZB, equivalent to generating 491 EB of data each day. The proliferation of Internet of Things infrastructure, smartphones, and wearable devices ensures that data generation is incessant. In this digital era, individuals have become fully digitized entities, heightening the urgency of efficient and accurate data processing. However, when the dataset to be searched is vast, the size of the corresponding index becomes substantial. Loading such a massive index from disk into memory can significantly elongate retrieval response times, impacting user search experiences.

The advent of big data also poses challenges to existing data management techniques. The predominant strategy for storing extensive data using databases involves three-tier memory-based parallel storage and querying. However, these approaches are marred by substantial hardware costs and necessitate the development of specialized database systems for management. The basic concept involves expanding hardware resources to accommodate greater storage capacity, albeit at the cost of increased query processing times. Although parallel database technology exploits multiple processors to achieve swift processing speeds, this comes at the expense of heightened hardware costs. Confronted with this immense data volume, elevating data transmission, retrieval, and computational speeds can facilitate the processing of more data streams within the same timeframe. This translates to swiftly accessing required data and making more efficient use of computational resources. Moreover, this approach enables the storage of more data within the same environment, resulting in shorter response times and optimized user experiences. Utilizing sliding windows for data compression conserves valuable memory resources. Given the high real-time demands of queries on data streams, the compression algorithms for data streams must not only possess high compression ratios but also the capability for swift compression and decompression. This not only enhances economic efficiency but also improves load capacity. The problem addressed in this paper is whether we can achieve a lower compression ratio while maintaining similar or superior compression performance compared to state-of-the-art lossy compression techniques.

In this paper, we propose a lossy compression framework based on the perceptron model for data prediction. By enhancing predictive accuracy, this framework attains stable high compression quality. Specifically, our contributions are as follows:

\begin{itemize}
    \item Introducing positive and negative factors based on the conventional relative-to-absolute domain transformation algorithm, thereby eliminating the need for additional symbol storage.
    \item Employing a three-layer perceptron as the predictor for improved predictive accuracy.
    \item Modifying the data selection rules for predictive surfaces to enable parallelized compression within compression blocks, enhancing compression speed.
    \item Compared to SZ2.1's PW\_REL mode, we achieved a maximum compression ratio reduction of $17.78\%$.
\end{itemize}

The remainder of this paper is organized as follows. In Section 2, we discuss related work. Section 3 outlines the optimization of mapping algorithms in the relative and absolute domains. Section 4 presents the design of the new compression framework. Section 5 evaluates our proposed solution. Finally, in Section 6, we provide a prospective outlook on future work.

\section{Related Work}
As the influx of vast amounts of data continues, the digitized society faces the challenge of efficiently managing such massive datasets. During data processing, the limitation of memory capacity may prevent the entire dataset from being stored in memory for subsequent computations. Moreover, the slow transfer speed between hard disk and memory introduces significant additional overhead due to multiple data transfers.

In response to these challenges, a range of lossless compression techniques has been developed, including Gzip\cite{ref9}, FPC\cite{ref10}, BlosC\cite{ref11}, and other such lossless compressors\cite{ref12}. Despite the widespread adoption of lossless compression techniques, their efficacy for scientific data remains limited. This limitation arises because lossless compression methods rely on repeated byte sequences, whereas scientific data often consists of diverse floating-point arrays. Consequently, research into lossy compression of scientific data has been ongoing for many years.

In our work, we have opted for a prediction-based model, as SZ has been recognized as a leading compressor in the realm of scientific data compression. In fact, the study of enhancing compression quality using SZ has spanned more than five years. SZ, originating from the Lorenzo predictor, introduced a comprehensive compression and decompression process\cite{ref1}. Subsequently, Di et al. proposed mapping multidimensional data to one dimension and predicting via optimal curve fitting\cite{ref2}. Tao et al. extended the one-dimensional Lorenzo predictor to the multidimensional case, broadening SZ's application scope. Furthermore, Tao et al. proposed the integration of SZ and ZFP to achieve improved compression effects through selection\cite{ref8}. Liang et al.\cite{ref4} introduced three predictors: the classical Lorenzo predictor, the average Lorenzo predictor, and the linear regression predictor factor. They evaluated which predictor performed better to select the most suitable one. Zhao et al.\cite{ref6} introduced second-order regression prediction and parameter selection methods to enhance predictive accuracy. Subsequently, they proposed dynamic cubic spline interpolation for prediction\cite{ref7}. During the developmental stage of lossy compression, Liang\cite{ref5} introduced a data transformation scheme, converting the pointwise relative error estimation compression problem into an absolute error estimation compression problem, making relative error an alternative criterion for compression selection.

\section{Optimization of Relative and Absolute Error Conversion}
In this section, we have made specific modifications to the original data transformation approach by incorporating two constant factors, thus eliminating the necessity of individual symbol recording for compression. Consequently, the optimized methodology ensures that no supplementary space is squandered on symbol storage during the compression procedure, thereby resulting in a further enhancement of the compression ratio. Subsequently, we commence by demonstrating that multiplication of the initial mapping scheme by a constant factor has no impact on the validity of the mapping formula. Subsequent to this, we undertake the derivation and substantiation of the numerical values of the constant factors.

\subsection{Theoretical support for optimization}
The crux of our inquiry lies in the quest for a factor that can be employed during the mapping procedure to effectively differentiate between positive and negative values. By doing so, the data inherently encapsulates the sign information, obviating any necessity for supplementary symbol storage space.

The following conclusion can be drawn: Multiplying the data in the relative domain by a constant factor does not exert any influence on the variation of the error bound.

We have the original formula:
\begin{equation}
    b_r=\frac{f^{-1}(f(x)+g(b_r))-x}{x}
\end{equation}

In the context of mathematical formulation, where $b\_r$ denotes the relative error bound, and $g(x)$ represents the function that maps the relative error bound to the absolute error bound.
\begin{figure}[!t]
    \centering
    \includegraphics[width=2.5in]{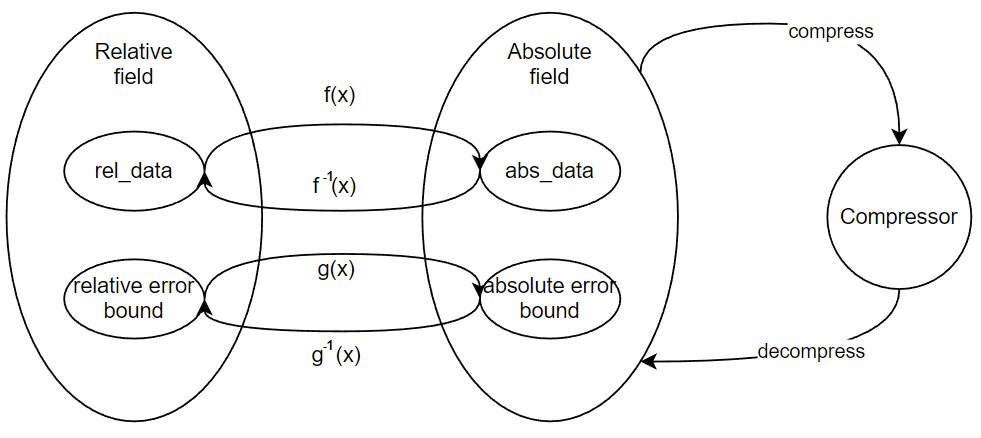}
    \caption{This is a schematic of how the mapping equation builds a link between the relative domain as well as the absolute domain.}
    \label{fig1}
\end{figure}

the resulting expression after rearrangement is as follows:
\begin{equation}
    f(x)+g(b_r)=f((1+b_r)x)
\end{equation}

In the preceding research, the mapping formula has been determined as $f(x)=\log_{base}x$, where $f(x)$ represents the outcome mapped to the absolute domain, and $x$ represents the data in the relative domain. It is important to note that the logarithmic function is uniquely employed as the mapping function, and the base basebase can be arbitrarily selected, adhering to the specified conditions. Accordingly, we now proceed with the expansion of $f(x)$.
\begin{equation}
    \log_{base}x+g(b_r)=\log_{base}(1+b_r)x
\end{equation}

At this juncture, we perform the operation of adding $$log_{base}C$$ to both sides of the equation, with *C* denoting an arbitrary constant.
\begin{equation}
    \log_{base}x+g(b_r)+\log_{base}C=\log_{base}(1+b_r)x+\log_{base}C
\end{equation}

Following that, the equation is reorganized to its original form $f(x)$.
\begin{equation}
    \begin{split}
        f(Cx)+g(b_r)=f((1+b_r)Cx)  \\
        \frac{f^{-1}(f(Cx)+g(b_r))-Cx}{Cx} = b_r
    \end{split}
\end{equation}

Through a comparative analysis with the original formula, we ascertain that pre-multiplying the data by a constant factor before the mapping operation does not compromise the accuracy and validity of the mapping process.

\subsection{The value of the factor}
We need two factors to act separately on positive and negative numbers, ensuring that positive values lie below the boundary hyperplane, and negative values lie above it. In this way, the problem can be formulated as finding suitable factors that result in negative numbers being greater than positive numbers after mapping.
\begin{figure}[!t]
    \centering
    \includegraphics[width=2.5in]{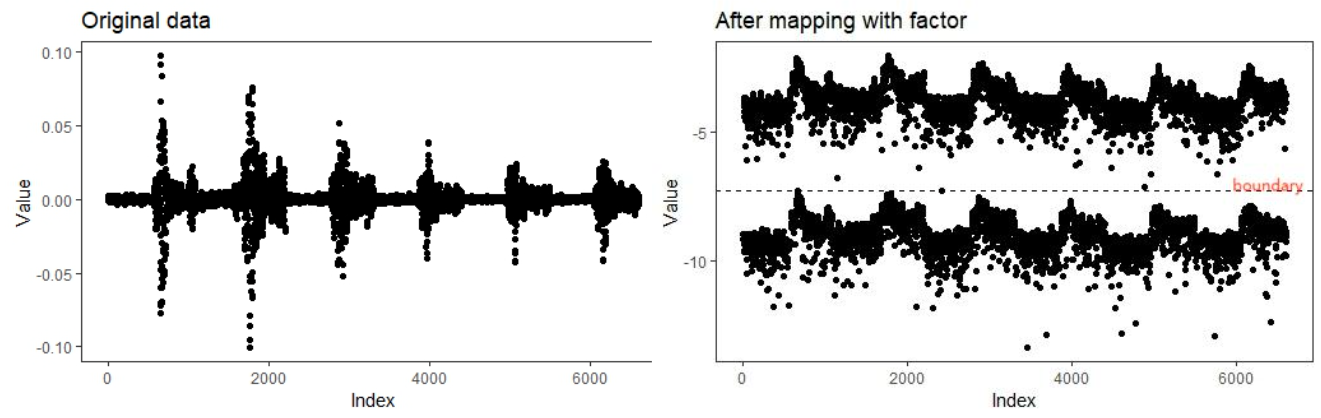}
    \caption{This is the distribution of the data after factoring, with a clear boundary between positive and negative numbers.}
    \label{fig2}
\end{figure}

Conclusions:

For positive numbers: 
\begin{equation}
    value_{abs}=\log_{base}(value_{rel}\times factor_{pos})
\end{equation}

For negative numbers: 
\begin{equation}
    value_{abs}=\log_{base}(value_{rel}\times (-factor_{neg}))
\end{equation}

For the error bound: 
\begin{equation}
    b_a=\log_{base}(1+b_r)
\end{equation}

Where:
\begin{equation}
    factor_{pos} = \left| abs\_min\right|
\end{equation}
\begin{equation}    
        factor_{neg} =\frac{((abs\_max+\epsilon_0)\cdot factor_{pos})\cdot (1+eb_{rel})^2}{\left| abs\_min\right| -\epsilon_0}
\end{equation}

Proof: We need to utilize two factors in the absolute domain to distinguish between positive and negative numbers. This can be expressed as follows:
$
    f((\left|abs\_min\right|-\epsilon_0)\times(factor_{neg}))-g(eb_{rel})>f((abs\_max+\epsilon_0)\times factor_{pos})+g(eb_{rel})
$

Here, {\bf{min}} represents the non-zero minimum absolute value of all data, and {\bf{max}} represents the maximum value of all data. We introduce {\bf{$\epsilon_0$}} to account for the rounding errors when storing data in computers, ensuring all data remains within controlled bounds.

Upon substitution of $f(x)$ and $g(x)$, our factors need to satisfy the following requirement:
$
(\left|abs\_min\right|-\epsilon_0)\times(factor_{neg})>(abs\_max+\epsilon_0)\times factor_{pos} \times (1+eb_{rel})^2
$

To ensure $factor_{neg}$ and $factor_{pos}$ do not exceed the range representable by a computer, we set$factor_{pos}=\left|min \right|$. This ensures that $factor_{pos}$ is of the same order of magnitude as the minimum data value. Substituting this into the equation (n), we obtain:
\begin{equation}
    factor_{neg}=\frac{(abs\_max+\epsilon_0)\times factor_{pos}\times (1+eb_{rel})^2}{\left|abs\_min \right|-\epsilon_0}
\end{equation}

By removing constant terms from this equation and substituting numerical values, we find that $factor_{neg}\propto abs\_max$. Thus,  $factor_{pos}$ and $abs_{max}$ are of the same order of magnitude, ensuring both factors remain within the computable range.

Finally, for data containing zeros, our approach is to treat them as rounding errors in computer representation, equivalent to the minimum representable floating-point number:
\begin{equation}
    value_{0_{abs}}=f(\epsilon_0\times factor_{pos})
\end{equation}

When performing the inverse mapping, if the data is less than or equal to $f(\epsilon_0\times factor_{pos})$, it is directly mapped back to 0.

Considering the presence of errors resulting from lossy compression, the actual absolute domain range corresponding to the value of 0 is $f(\epsilon_0\times factor_{pos})\pm g(eb_{rel})$. To ensure proper mapping, we compare this range with the boundary $boundary=f((abs\_max+\epsilon_0)\times factor_{pos})+g(eb_{rel})$ Therefore, data containing 0 will always be inversely mapped according to the positive number rule.When the value of 0 is mapped back to the relative domain, its data range is $\left \{ 0 \right \} \cup(\epsilon_0,\epsilon_0\times(1+eb_{rel})]$, which is acceptable.

\section{Overall framework}
Within this section, our initial focus lies in retrospectively examining the existing architecture of SZ. Subsequently, we conduct a thorough analysis of the challenges intrinsic to the SZ compression mechanism. An in-depth understanding of these challenges is of paramount importance for appreciating the profound significance of our proposed resolution in substantially augmenting compression efficiency.

\subsection{A Review of the SZ Lossy Compression Framework}
Based on the research and application efforts of numerous scholars, the SZ compressor has been widely acknowledged as an exemplary lossy compression tool.

The SZ compressor encompasses four distinct stages:
\begin{itemize}
    \item Prediction: In the prediction phase, SZ forecasts the current data by utilizing neighboring points, employing various prediction methods in different versions.
    \item Linear Scale Quantization: In this phase, SZ computes the prediction error between the forecasted outcome from the previous stage and the actual data. This prediction error is then linearly quantized into integers based on the absolute error bound, producing quantization codes.
    \item Huffman Coding: SZ subsequently transforms the quantization codes into Huffman codes using a tailored Huffman coding scheme.
    \item Lossless Compression: Lastly, SZ employs a lossless compression algorithm to further enhance the compression ratio of the SZ compressor.
\end{itemize}

Firstly, SZ is a highly flexible compression framework, where data prediction is the most critical step. More accurate prediction will lead to the concentration of prediction errors of data tending to zero, which can be quantified in a smaller range and compressed with Huffman coding using fewer bits, resulting in better compression ratio. Therefore, how to optimize the prediction process so that any sequence can be transformed into a small-range sequence after prediction will be the focus of the next step.

At the heart of SZ's prediction mechanism lies the utilization of previously forecasted neighboring points to predict the current target point. Its core formula is: $$F=\sum\limits^{(k_1,k_2)\neq(0,0)}_{0\leq k_1,k_2\leq n}(-1)^{k_1+k_2+1}\left(\begin{matrix}n \\ k_1\end{matrix}\right)\left(\begin{matrix}n \\ k_2\end{matrix}\right)V(k_1,k_2)$$, where $F$ represents the predicted value of the target point, and $k_1$,$k_2$ denote the neighboring forecasted points. If we interpret this convolution process as the output of a single-layer perceptron, optimizing the initial values and refining the network structure can to some extent reduce the prediction error and ultimately achieve better compression ratios. Thus, we have devised a three-layer perceptron for the prediction process within the compressor.

\subsection{Perceptron}
In the prediction phase, we opt to replace the original prediction formula with a perceptron. To enhance the model's performance without introducing excessive computational overhead, we construct a perceptron. The presence of hidden layers enables the model to better capture non-linear characteristics. However, considering that overly complex models may lead to additional computational time, we choose a three-layer perceptron. 

In the input layer, the number of neurons is based on the number of data points on the prediction surface of the SZ predictor (as exemplified in this paper with a one-layer Lorenzo predictor), with an additional bias neuron. In the hidden layer, to avoid excessive computational complexity, we set it to contain two neurons. Subsequently, we introduce a leaky ReLU activation layer to enhance the model's predictive capabilities. We opt for the leaky ReLU activation layer because it is relatively simple, and Leaky ReLU maintains a non-zero derivative in the negative range, which can improve performance. Finally, the output layer consists of a single neuron. Since our objective is to predict scientific data, we refrain from adding additional activation layers in this context.
\begin{figure}[!t]
    \centering
    \includegraphics[width=2.5in]{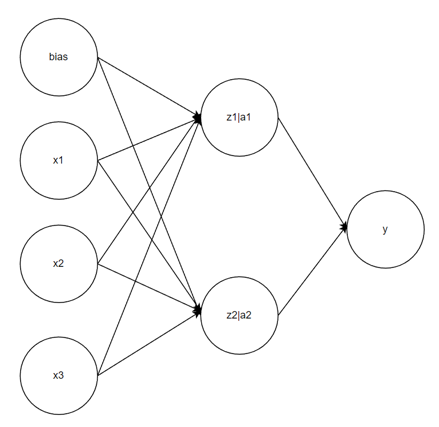}
    \caption{This is a schematic of the structure of a perceptron.}
    \label{fig3}
\end{figure}

In addressing the matter of model parameter initialization, we align the initialization process for model parameters spanning from the input layer to the hidden layer with the weights utilized within the SZ predictor. This alignment is represented by $(-1)^{k_1+k_2+1}\left(\begin{matrix}n \\ k_1\end{matrix}\right)\left(\begin{matrix}n \\ k_2\end{matrix}\right)$. Concurrently, we set the initialization of model parameters linked to bias neurons at a value of 0. When dealing with parameters connecting the hidden layer to the output layer, we initialize these parameters as two distinct numbers, their sum equating to 1. This configuration lends itself to a conceptualization of our model as a weighted amalgamation of two SZ predictors.

\subsection{Tranning}
Firstly, we divide the data to be compressed into blocks according to the requirements of the compression process. Subsequently, a representative training set is created by uniformly sampling from the blocks of data to enable the perceptron to achieve improved predictive performance. Empirical observations indicate that a uniform sampling rate of 10% exhibits stability, significantly reducing the training set size without substantial loss of data features.

Subsequently, the training dataset is partitioned into a training set and a validation set at an 8:2 ratio. Given that L2 regularization is incorporated during the training phase to enhance model generalization, multiple training rounds are not necessary due to the similarity between data blocks. One training round effectively encompasses training a data block through multiple rounds. Therefore, one training round yields satisfactory outcomes. For hyperparameter selection, encompassing learning rate and regularization parameters, to avoid repetitive parameter tuning during training, we concurrently train multiple sets of hyperparameters and ultimately select the optimal outcome from the best-performing group.

Next, regarding the selection of the range of quantization codes, we recognize that quantization codes span from $0$ to $2^m-1$. When the value of $m$ is uncertain, we first choose an initial value of $m$ based on the characteristics of the data to be compressed. Following the standard compression and decompression processes, and upon achieving relative error values in accordance with user-defined thresholds, we introduce the coverage metric to determine the value of $m$. The coverage rate signifies the proportion of data covered by the quantization codes relative to the entire dataset, denoted as $Coverage=\frac{covered \ data}{all \ data}$. Through experimentation, it has been determined that when the relative error is less than 1.5%, $m$ should be chosen for a coverage rate closest to 50%, while for relative errors exceeding 1.5%, $m$ should be selected to yield a coverage rate closest to 80%.

Furthermore, during the validation phase, we need to determine the optimal model performance, which involves simulating the real compression process on the validation set. Following the compression procedure step-by-step, including prediction, quantization, Huffman coding, and writing to a bit stream, the current compression ratio is calculated. The decompression process is then performed to compute the current relative error. A model is considered effective if the actual relative error remains within 1.5 times the user-defined threshold, and it exhibits a lower compression ratio during this process.

Simultaneously, after obtaining an appropriate model and selecting $m$ during the validation phase, we need to compute the frequency of each quantization code in each block of the validation set. This information is then used to derive the average frequency of each quantization code per block, forming the basis for constructing a Huffman tree and generating corresponding Huffman codes. These codes are retained for direct retrieval during the compression phase. Consequently, the Huffman codes employed during compression are not optimal solutions for each block but are derived as suboptimal solutions based on the validation set. This approach saves considerable time otherwise spent on constructing Huffman trees.

\begin{figure}[!t]
    \centering
    \includegraphics[width=2.5in]{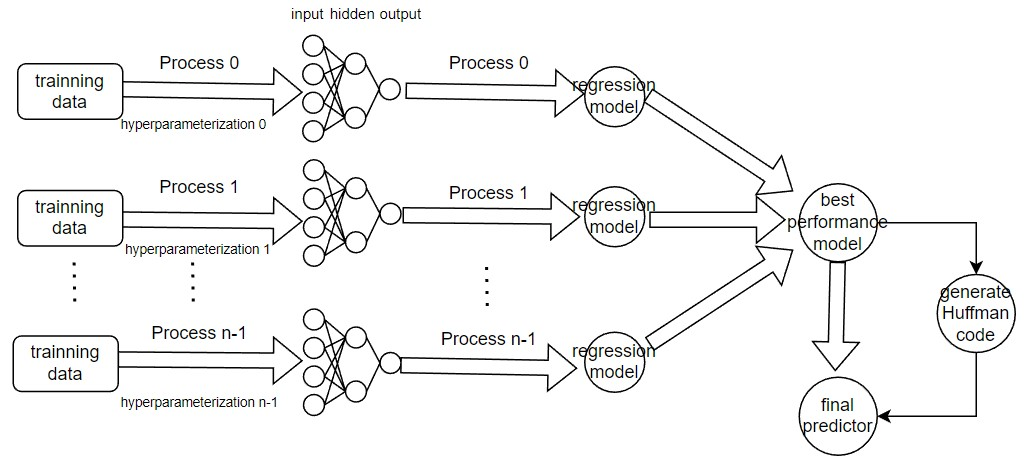}
    \caption{This is the structure of a parallel multi-hyperparameter training.}
    \label{fig4}
\end{figure}

\subsection{Compressed}
Within the compression stage, the initial data is employed as the input and subjected to prediction through the trained model, yielding prediction deviations. These deviations are sequentially subjected to quantization coding and Huffman coding processes, culminating in the encoding being written into a bitstream.

\subsubsection{Selection of Predictive Surface Data}
\ 
\newline \indent During the prediction phase, it is noted that the SZ compressor utilizes prior prediction data as the input for subsequent predictions. This methodology effectively maintains consistency in the decompression stage, thereby ensuring that data remains entirely within the scope of the error range. However, it has been observed that such predictions often progressively deviate from the original data. We postulate that employing authentic values of the data slated for compression as input could yield enhanced predictive outcomes, consequently boosting the overarching compression ratio. However, this paradigm introduces a quandary during the decompression phase, wherein the decompressed data must similarly serve as prediction input. Yet, the compressed data is inevitably tainted by errors, and the persistent utilization of flawed data as model input would inexorably culminate in error accrual, potentially leading to unmanageable ramifications.

To circumvent the predicament of cumulative errors, we propose two methodologies.
\paragraph{Suitable regularization parameters}
\ 
\newline \indent Throughout the training phase, we engage in parallelized training featuring multiple regularization parameters, aiming to acquire models showcasing divergent generalization capacities. By subjecting the models to straightforward decompression assessments on the validation dataset, we identify the minimal regularization parameter that forestalls excessive error accumulation while simultaneously aligning with the stipulated user-defined error thresholds. This course of action ensures that predictive accuracy is preserved to a reasonable extent while adhering to the user-prescribed error tolerance.
\paragraph{Equilibrium Points}
\ 
\newline \indent During the compression phase, instances of data that extend beyond the quantization range are frequently generated, subsequently conserved as uncompressed data. In the course of decompression, whenever a marker signifying uncompressed data is encountered, data is directly retrieved from the uncompressed data stream. Consequently, this uncompressed data remains unblemished by errors and is denoted as equilibrium points. The presence of equilibrium point mechanisms fosters a tendency toward error convergence during the prediction phase of decompression, specifically when equilibrium points are manifest on the predictive surface. Thus, by harnessing equilibrium points, the compressor can exercise control over errors within a delimited span, either adhering to or marginally surpassing the user-specified relative error prerequisites.

Given that a certain quantity of equilibrium points inherently exists within the compression phase, our primary strategy involves adjusting regularization parameters to manage errors while optimizing the compression ratio. Only when identifying suitable regularization parameters becomes infeasible do we resort to deliberately introducing a defined number of equilibrium points into naturally sparse regions of equilibrium points, aiming to achieve targeted error control.

\subsubsection{Parallelized Prediction}
\ 
\newline \indent In previous SZ research, concurrent prediction during data parallelism was unattainable due to the necessity of knowing the predicted values of neighboring points. However, to enhance prediction accuracy, we ensured that the data on the prediction surface came from the actual values of previously predicted data, thereby enabling parallel prediction. During the compression phase, only the actual values of the data to be compressed were needed. However, during the decompression phase, due to the presence of cumulative errors, we remained uncertain about the real values of the points the current predicted point depended on, necessitating a serial execution for decompression. In contrast, during the compression phase, since all the points to be predicted were based on actual values, we could directly access these values. This allowed us to initiate prediction from any point without affecting the overall prediction results. Hence, we have demonstrated the reliability of the parallel prediction approach. We divided the data blocks to be compressed into static tasks for each thread, ensuring that different threads could predict simultaneously. Adhering to the principles of load balancing and locality, we employed a row-wise static partitioning within the data blocks. Consequently, we set the starting task point for each thread as the balance point, guaranteeing that during decompression, each sub-block could commence decompression from the balance point, thereby controlling the overall error.
\begin{figure}[!t]
    \centering
    \includegraphics[width=2.5in]{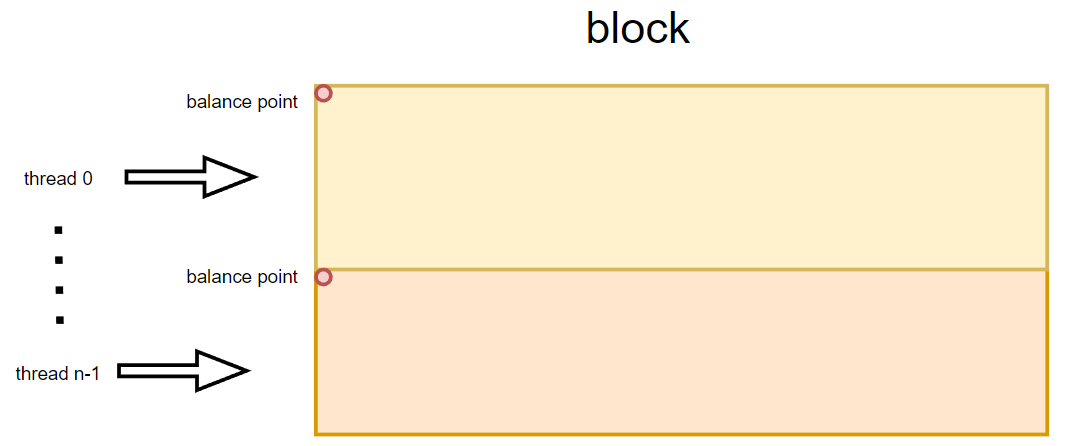}
    \caption{Multi-threaded simultaneous prediction from different starting points.}
    \label{fig5}
\end{figure}

\section{Implement}
In this section, we discussed how to effectively utilize data transformation schemes and perceptrons in the compression process.

The overall process, as shown in Figure 6, comprises two main components: the training phase and the compression phase.

For the training phase, we start by sampling the original data to form a training dataset. The more comprehensive the features in the training dataset, the better the prediction performance of the corresponding trained perceptron, thereby enhancing the overall compression ratio. Subsequently, we input the training dataset into the initialized perceptron and conduct parallel training with multiple sets of hyperparameters. During the training process, we also obtain the m-value mentioned in Section 4.3. Following this, from the multiple sets of models, we choose the group of models with the most effective performance and the corresponding Huffman codes. These two components will be applied in the actual compression process.

As for the compression process, we begin by passing the data to be compressed to the pre-trained model to obtain its predicted value. Next, we calculate the difference between the predicted value and the original value, forming the prediction error, which is then quantized. The range of quantization error should be calculated based on the m-value obtained during the training process. Data beyond the quantization range is referred to as balance points, and we save the original information of these data points and write a flag for balance point data in the bitstream. For data within the quantization range, we encode it into Huffman codes using the Huffman code key-value pairs obtained during the training phase and finally write it into the bitstream. The result of this process is the compressed data stream, i.e., the bitstream, along with the balance point dataset, and the associated $factor_{pos}$, $factor_{neg}$ for this data block. These data will be utilized in the decompression process.

During the decompression process, we reconstruct the Huffman tree based on the Huffman code key-value pairs. Then, we read bit by bit from the bitstream and decode accordingly using the Huffman tree to obtain the corresponding quantization code. If the read quantization code is a flag for balance point data, we directly retrieve the corresponding data from the balance point set. If it is not a flag for balance point data, we restore its prediction error in accordance with the inverse process of the quantization process. Simultaneously, our perceptron will traverse from the beginning of the data and use the prediction error for correction to ensure the accuracy of the current predicted data and the accuracy of predicting the next data bit.

The pseudocode for lossy compression based on perceptrons is presented in Algorithm 1.
\begin{figure}[!t]
    \centering
    \includegraphics[width=2.5in]{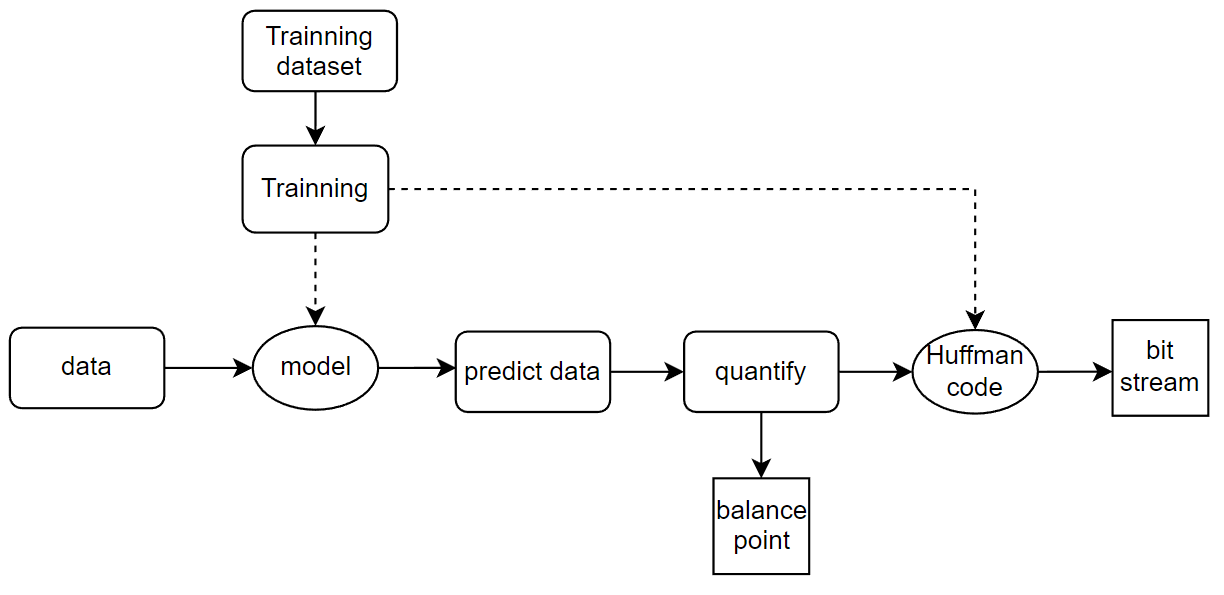}
    \caption{Compression Ratios of the Two Compressors at Different Scale Relative Errors.}
    \label{fig6}
\end{figure}

\begin{algorithm}[h]
	\caption{Lossy compression based on perceptual machines}%算法标题
    \renewcommand{\algorithmicrequire}{\textbf{Input:}}
    \renewcommand{\algorithmicensure}{\textbf{Output:}}
	\begin{algorithmic}[1]%一行一个标行号
		\REQUIRE a dataset(denoted by $D$), the max/min number in dataset(denoted by $max_D/min_D$),user-specified point-wise relative error bound $b_r$.
        \ENSURE compressed data stream in form of bytes and balance point dataset.
        \STATE $b_a^{\prime} = log_2(1+b_r)$;
        \STATE $factor_{pos} = \left| min_D\right|$;
        \STATE $factor_{neg} = \frac{(abs\_max+\epsilon_0)\times factor_{pos}\times (1+eb_{rel})^2}{\left|abs\_min \right|-\epsilon_0}$;
        \FOR{each data poin $D_i$ in the dataset $D$}
            \IF{$D_i == 0$}
                \STATE $d_i = log_2(\epsilon_0 \times factor_{pos})$;
            \ELSE 
                \IF{$D_i>0$}
                    \STATE Compute $d_i=log_2(D_i\times factor_{pos})$;
                \ELSE 
                    \STATE Compute $d_i=log_2(D_i\times factor_{neg})$;
                \ENDIF
            \ENDIF
            \STATE Pass the data point proximity data as input to the three-layer perceptron and get the output $d_i^{\prime}$
            \STATE Compute $error = d_i^{\prime} - d_i$;
            \STATE Quantize the error to get the corresponding quantization code q
            \IF{q!=0}
            \STATE Encoding q into a Hoffman code based on the Hoffman code obtained in the training phase h
            \STATE Write h to the bit stream
            \ELSE \STATE Write the original data $d_i$ to the set of equilibrium points
            \ENDIF
        \ENDFOR
        \STATE Output the compressed data stream in bytes and uncompressed data;
    \end{algorithmic}
\end{algorithm}

In the \textbf{algorithm 1}, we first calculate the required absolute error bound, denoted as $b_a'$, based on the user-defined relative error bound $b_r$. Simultaneously, we compute the corresponding values of $factor_{pos}$ and $factor_{neg}$ based on the maximum and minimum values of the input data. Subsequently, we perform data transformation for each individual data point (lines 5 to 13). Following the data transformation process, we proceed with the lossy compression process, specifically tailored to the absolute error bounds (lines 14 to 21). During this process, we predict the data and obtain prediction errors. These prediction errors are then quantized. For data points falling outside the quantization range (denoted as $q==0$), we treat them as equilibrium points, preserving the original data, and write the quantization code into the bitstream. For data points within the quantization range, we retrieve their Huffman codes from a lookup table and write the Huffman codes into the bitstream. Finally, the compressed bitstream and the equilibrium point dataset are output.

\section{Evaluation}
\subsection{Experimental Setup}
\subsubsection{Execution Environment}
We prepared two experimental environments for our study. The first experimental environment involved a personal computer equipped with an 8-core AMD processor and 14GB of memory, and the files were accessed using traditional single-file sequential access methods. %In the second experimental environment, 

The application data primarily consisted of seismic data, and our evaluations were focused on datasets of substantial scale. This approach better highlights the significance of directed optimization for prediction capabilities through pre-training.

\subsubsection{dataset}
\ 
\newline \indent SGY\@: SGY file is a data file saved in the SEG-Y (Society of Exploration Geophysicists) format. It contains geophysical data in binary and textual format, which includes the coordinates of reflected seismic waves. SGY files may store sweep frequency, types, and length, impulse signal polarity, projection zone and method, and other metadata.

\subsubsection{Evaluation Metrics}
\begin{itemize}
    \item Compression rate (CR) based on the same error bound: $$\frac{conpressed\ size}{original\ size}\times 100\%$$
    \item Compression speed and decompression speed: $$\frac{original\ size}{compression time}(MB/s)$$ and $$\frac{reconstructed\ size}{decompression\ time}(MB/s)$$ 
\end{itemize}

\subsection{Evaluation Results and Analysis}
We will compare the performance of our proposed solution with the existing SZ2.1, which is capable of achieving relative error-constrained compression.
\begin{figure}[!t]
    \centering
    \includegraphics[width=2.5in]{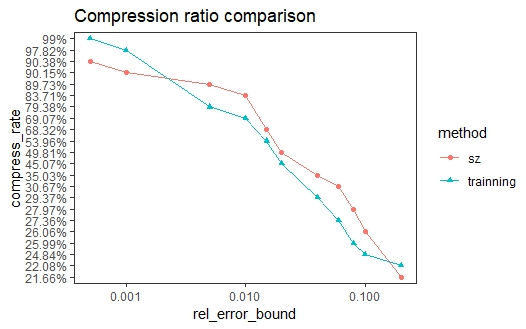}
    \caption{Compression Ratios of the Two Compressors at Different Scale Relative Errors.}
    \label{fig7}
\end{figure}

We observe two intersections in the graph. Below a relative error of 0.2%, although our proposed solution achieves a higher compression ratio than SZ, the compression effects of both methods are not significant. However, at relative errors exceeding 20%, due to the requirement for a certain amount of equilibrium points to prevent excessive error accumulation, our solution's compression ratio remains higher than SZ. Notably, within the intermediate relative error range of 0.2% to 20%, our solution consistently outperforms SZ.

\begin{figure}[!t]
    \centering
    \includegraphics[width=2.5in]{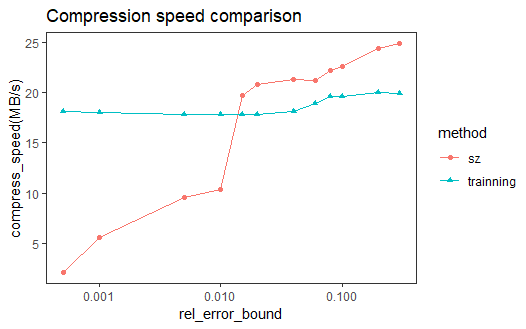}
    \caption{Compression Speeds of the Two Compressors at Different Scale Relative Errors.}
    \label{fig8}
\end{figure}

Regarding compression speed, we compare the compression speeds under different relative error conditions. Employing an intra-block parallel prediction mechanism while having a more complex neural network structure than SZ leads to a compression speed closely aligned with SZ. However, due to the presence of the intra-block parallelization mechanism, our proposed compressor, given sufficient hardware resources, can ultimately outpace the original SZ compressor in terms of speed.

Lastly, in terms of accuracy, since compression accuracy relies on the number of regularization parameters and equilibrium points, the aforementioned data measurements are based on meeting user-defined accuracy. If users require higher accuracy, they can adjust regularization parameters or the number of equilibrium points to achieve extremely high accuracy, albeit at the expense of compression ratio and speed performance. Therefore, conducting additional measurements in this context is unnecessary.

\section{Conclusion and Future Work}
In this article, we have modified the existing mapping formula from relative domain to absolute domain. We have introduced a novel framework utilizing a neural network as the predictor and adjusted the selection of prediction surface points to introduce an intra-block parallel structure. Comparisons were conducted with the SZ2.1 version in its PW\_REL mode to estimate the compression ratio and performance of our proposed solution. The main findings can be summarized as follows:
\begin{itemize}
    \item Our analysis indicates that the neural network predictor's complex structure introduces non-negligible additional time overhead.
    \item Compared to SZ's relative error mode, our solution achieves an improvement in compression ratio at commonly used relative error scales, with a maximum enhancement of 17.78\%.
    \item Our solution exhibits similar compression performance to SZ2.1, while also possessing an intra-block parallel structure, thereby having higher performance enhancement potential than SZ.
\end{itemize}

In future work, we plan to enhance compression quality further by optimizing the neural network prediction model, as well as boosting performance through code optimization implementations.

%\section{References}

\begin{IEEEbiographynophoto}{Jane Doe}
Biography text here without a photo.
\end{IEEEbiographynophoto}

% \begin{IEEEbiography}[{\includegraphics[width=1in,height=1.25in,clip,keepaspectratio]{./photo/profile2.png}}]{IEEE Publications Technology Team}
% In this paragraph you can place your educational, professional background and research and other interests.\end{IEEEbiography}

\end{document}